# The Thirty Meter Telescope International Observatory facilitating transformative astrophysical science

## Warren Skidmore[1,*], G. C. Anupama[2] and Raghunathan Srianand[3]

[1]Thirty Meter Telescope International Observatory, 100 W. Walnut St., Suite 300, Pasadena, CA 91124, USA
[2]Indian Institute of Astrophysics, 2nd Block, Koramangala, Bengaluru 560 034, India
[3]Inter-University Center for Astronomy & Astrophysics, Post Bag 4, Ganeshkhind, Pune 411 007, India

**The next major advancement in astronomy and cosmology will be driven by deep observations using very sensitive telescopes with high spatial and spectral resolution capabilities. An international consortium of astronomers, including Indian astronomers are building the Thirty Meter Telescope to achieve breakthroughs in different areas of astronomy starting from studies of the solar system to that of the early universe. This article provides a brief overview of the telescope, science objectives and details of the first light instruments.**

## Introduction

THE Thirty Meter Telescope (TMT) observatory is being constructed by an international partnership in order to tackle leading questions in astrophysics, see Figure 1. The partnership collectively brings together the necessary resources in terms of engineering skills and manufacturing capabilities needed to design, construct and operate what is arguably the most ambitious ground-based optical/infrared observatory ever constructed, and India is a leading partner in the project.

Other partners are Japan, China, Canada, the University of California and Caltech in USA, with funding also from the Gordon and Betty Moore Foundation. The partnership was first formed in 2004 and now all partners are contributing significant resources to the project with responsibilities for delivering subsystems and instruments shared between all parties.

India specifically has responsibility for the development of the observatory software, the adaptive optics guide star catalogue, the primary mirror support and control system hardware, primary mirror segment polishing and various aspects of instrument development. Indian astronomers and engineers are also involved collaboratively with other aspects of the observatory development such as systems engineering, the telescope optics alignment and phasing system (APS) and scientific leadership. Indian scientists are involved with the TMT International Science Development Teams who are developing science cases that are helping to define the observatory operations, guide the first light instrument development and inform future instrument choices.

## The driving scientific questions

The main questions in astrophysics at the present time that cannot be effectively answered with existing astronomical facilities and therefore require the construction of new facilities are discussed below.

*What is the nature of dark energy and dark matter?*

The evolution rates of cosmic distances and cosmic structures are dependent on the properties of dark energy. Highly sensitive, high spatial resolution spectroscopy of distant supernovae is key to constraining the properties of dark energy. The accelerating expansion of the universe was discovered by measuring the distance to high-redshift type-Ia supernovae and their recession velocity. However, there is much more to be done to improve and expand this mode of studying the impact of dark energy.

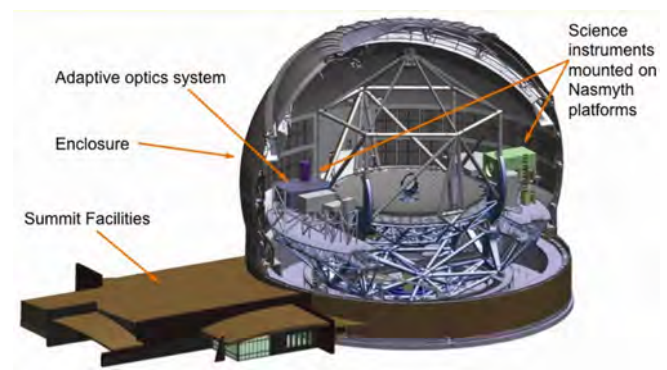

**Figure 1.** An overview of the design for the Thirty Meter Telescope Observatory.

---

*For correspondence. (e-mail: was@tmt.org)

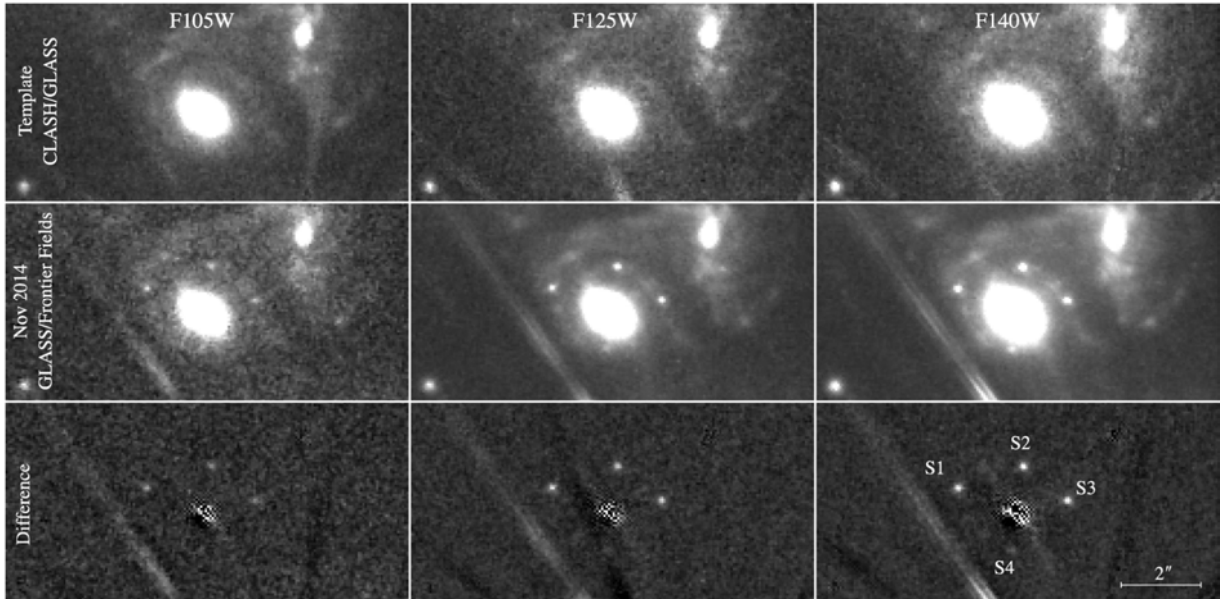

**Figure 2.** HST WFC3 images showing a supernova gravitationally lensed by an early type galaxy cluster[3].

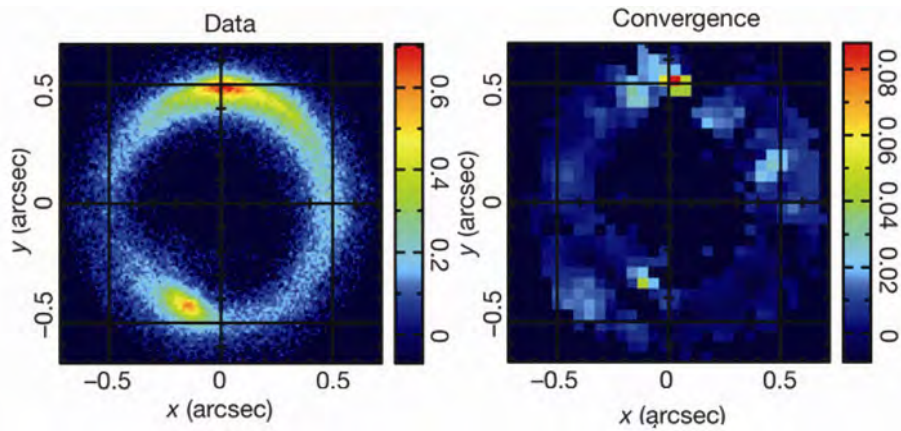

**Figure 3.** (Left) Keck/NIRC2 imaging of the B1938+666 system[4], with the lensing galaxy masked out. (Right) Enhanced peak in the convergence map derived from fitting the data corresponding to the location of the substructure in this system.

Time-delay cosmography is the measurement of the time delay for multiple gravitationally lensed images of background supernovae by foreground galaxies and provides a means to constrain the dark energy equation of state[1–3] (see Figure 2).

Variations in the dark matter distribution on the smallest spatial scales depend on the properties of the dark matter particle, i.e. warm versus cold particles, how the dark matter particles interact with themselves and with ordinary baryonic matter. Higher spatial resolution imaging of gravitationally lensed background galaxies will reveal the small anomalies in the lensed images and expose the dark matter distribution within the lens (Figure 3)[4].

Measurements of the 3D velocities of individually resolved stars in the cores of nearby dwarf galaxies using high spatial resolution spectroscopy and high accuracy astrometry will let us determine the level of compactness of the dark matter halo down into the centre of the dwarf galaxies.

*Evolution of fundamental physical 'constants'*

There are several arguments which suggest that what are considered fundamental physical constants could, or even should, vary over cosmic time. Such arguments arise in string theories and higher dimensional cosmological theories[5]. Searching for evidence for the space and time variations of the physical constants is an extensive activity amongst both experimental physicists and astronomers.

High signal-to-noise, high spectral resolution observations of absorption lines in the spectra of distant quasars

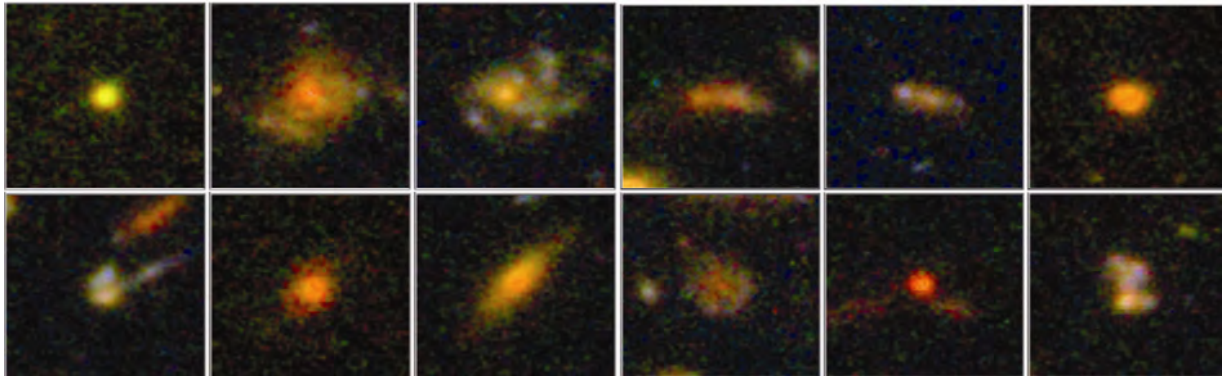

**Figure 4.** Set of images of galaxies between redshift 1.4 and 3, from the HST CANDELS survey[8,9], showing diverse morphologies, including highly compact objects, dust-enshrouded disks and highly disturbed merger candidates. With IRIS on the TMT it will be possible to observe systems of this nature with unprecedented angular resolution in the rest-frame visible.

($z > 2.5$) provide a means to probe changes in the fine-structure constant $\alpha$ by measuring the wavelengths of sets of absorption lines of heavy elements arising from transitions with different sensitivities to $\alpha$. One can also use HD and $H_2$ molecular absorption lines to place strong constraints on the ratio of electron-to-proton mass, $\mu$. Presently, the tight constraints of these parameters have uncertainties that are consistent with either evolving or constant values[6,7]. The TMT will give higher signal-to-noise measurements for more distant quasars and reduce the uncertainties by several factors.

Complementary studies can be carried out with the uGMRT, ALMA and SKA using radio line spectra to investigate other fundamental constants.

*When did the first galaxies form and how did they evolve?*

The first stars formed from the primordial material created in the Big Bang, before subsequent generations of stars had begun enriching the universe with the heavier elements that influence the nature of further star formation. The sensitivity and high spatial resolution of the TMT will for the first time allow one to characterize the star formation process and the early chemical enrichment process in the very first galaxies (Figure 4)[8,9].

Complementary to the spatially resolved studies of nearby dwarf galaxies mentioned above, determining the individual sets of stellar populations within each dwarf galaxy would be proof of the stochastic nature of galaxy assembly from smaller galaxies. This requires high spatial resolution observations. The TMT is designed to provide the required spatial resolution for this purpose.

*Chemical evolution of galaxies and its relation to star formation*

In order to understand how chemical enrichment is driven by stellar processes throughout cosmic time up to the present day in the nearby universe, a range of different studies is required. In the distant universe at $z \sim 6$, pair-instability supernovae of massive stars formed from low-metallicity material are important sources of enrichment. To study the composition of the ejecta requires high spectral resolution observations ($R \sim 20{,}000$ to $100{,}000$)[10] on extremely large telescopes to give the required signal-to-noise ($S/N > 50$ to $>100$). Molecules can form from the ejected metals and will be detectable with the TMT at redshifts of $z \sim 6$ along with other emissions from the partially enriched interstellar medium within early galaxies. Metallicity studies of homogenous populations reveal the isotope ratios and indicate the relative effects of stars following different stellar evolution tracks and the amount of recycling. The TMT with its wide spectral coverage and high spatial resolution will be ideally suited for this purpose.

*How are super-massive black holes related to the properties of their host galaxies?*

The reasons why we see a strong correlation between the mass of the central super-massive black hole (SMBH) and the properties of the central galactic bulge in nearby galaxies (Figure 5)[11], are not understood. The 10 m class telescopes do not have the spatial resolution and sensitivity to measure the SMBH mass and host galaxy properties at significant enough distances to study the evolution of the relation over cosmic time. The TMT with diffraction-limited adaptive optics will allow the SMBH mass and host galaxy properties to be measured to distances 20 times as far as is done today and to 10× lower SMBH mass in much smaller galaxies, increasing the number of galaxies that SMBH can be determined by a factor 1000 compared to present facilities.

The TMT will be able to study the properties of the first quasars to ever form, the earliest generation of accreting massive black holes. Such distant quasars will be discovered photometrically by the next generation of wide field survey telescopes such as LSST and Subaru with Hyper-Suprime-Cam, but the targets will be so faint that only 30 m class telescopes will have the sensitivity to spectroscopically characterize the processes taking place in the quasar central engine.

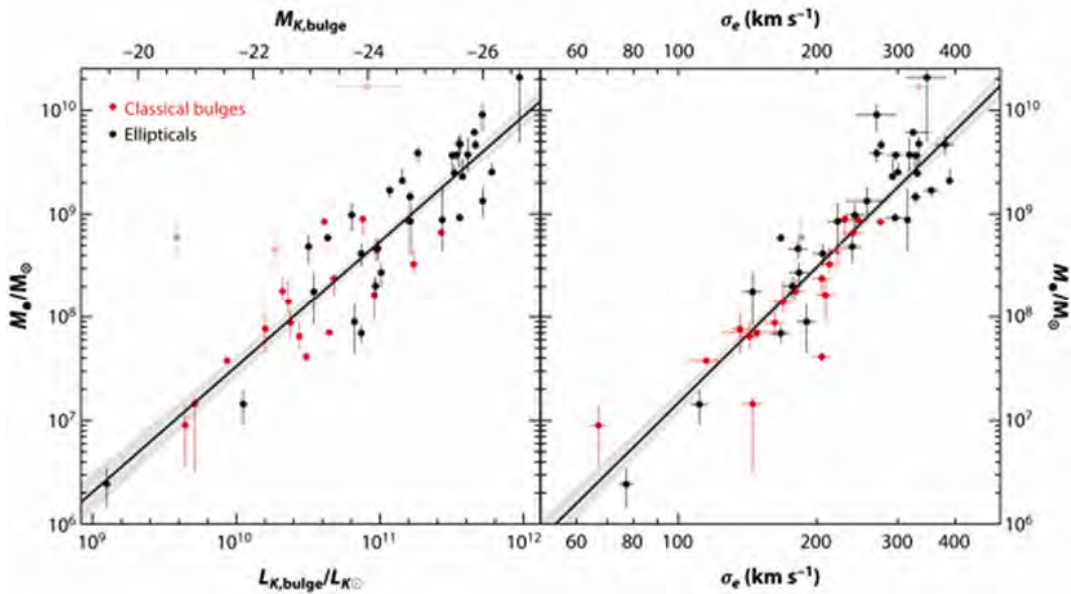

**Figure 5.** Correlation between black hole mass and bulge properties for nearby galaxies: (left) K-band luminosity and (right) stellar velocity dispersion[11].

The TMT will also be able to carry out intensive studies of the inner regions of moderate redshift quasars where the sphere of influence of some of the higher mass SMBHs can be spatially resolved, greatly advancing our understanding of the processes driving SMBH growth and influencing the evolution of their host galaxies.

*Mysterious transient objects, time critical and rapidly variable phenomena*

Massive stars are believed to die in energetic explosions as supernovae or gamma-ray bursts (GRBs). These highly luminous explosions are detectable at large cosmological distances with 8 m class survey telescopes and can be used as powerful probes of the environments and star formation processes within early galaxies when studied spectroscopically with the TMT. Observations must begin within a few minutes from the moment of detection in the case of GRBs and within about an hour in the case of supernovae in order to capture the evolving emission from the exploding shell. The early phase evolution will enable distinguishing the progenitors of these events and help with understanding the supernovae explosion mechanism, a key aspect for improving the accuracy of cosmological distance measurements with supernovae.

The high sensitivity of the TMT will make it possible to study processes that change very rapidly in objects like low-mass X-ray binaries, where accreting material causes flashing and flickering at timescales as short as fractions of a second, or pulsars with regular variations on spin periods that can be measured in milliseconds.

Most intriguing are the faint, completely unknown high-amplitude transient objects that are expected to be found in significant numbers in photometric surveys from 8 m class telescopes. The TMT will have the rapid response capabilities coupled with a suite of versatile instruments from the beginning of operations that will allow detailed studies of these enigmatic but potentially enlightening phenomena. A powerful synergy with the uGMRT and SKA would be the study of rapid radio transients which are extremely powerful, distant events that have durations of milliseconds, but have not been detected in the optical or infrared.

*What are the processes of star and planet formation?*

Improving our understanding of the star and planet formation processes is fundamental towards fully understanding how galaxy and chemical evolution affects the types of stars and planets that are formed and hence affects the prospects for the existence of life elsewhere in the universe. Basic aspects of the star formation process, such as what determines the stellar mass distribution, are unknown at this time[12].

The high spatial resolution of the TMT will allow for the first time, the mass distributions of stars to be measured in star-forming regions that are in the variety of different types of neighbouring galaxies outside of the Milky Way. This will mean that the star formation process within a wide range of different environments can be studied.

*What are the characteristics of extra-solar planets and is there life elsewhere in the universe?*

In order to better understand the range of planet types and to search for signatures of living processes on other planets, it is necessary to characterize the atmospheres of those planets, to identify the molecular species present within the atmospheres. The 10 m class telescopes have been used to conduct rudimentary studies of the atmospheres of a few of the closest, recently formed hot Jupiter planets[13]. Space telescopes have been able to coarsely map the surface temperature distribution of lava rock planets[14]. To be able to characterize low-mass, cool, mature, earth-like planets in the habitable zones of their host stars we need the high spatial resolution, high contrast and sensitivity provided by the TMT. High sensitivity is also needed in order to measure the tiny signal caused when a fraction of light of an exoplanet host star passes through the atmosphere of a transiting planet.

**Capabilities required for a next generation optical/infrared ground-based observatory**

*General description of observatory*

The observing programmes that result from the major scientific questions discussed above require a well-defined suite of instruments with new capabilities. Many of the science programmes require high sensitivity and high spatial resolution. The greatest advances with a given telescope occur when adaptive optics is used, allowing the fundamental diffraction-limited resolution of the telescope to be reached, and greatly enhancing the sensitivity relative to existing 8 m and 10 m observatories. When considering the major scientific questions, it is clear that a telescope with diameter of about 30 m and equipped with appropriate forms of adaptive optics systems will provide the required leap forward in capabilities when compared to existing 10 m class telescopes.

The TMT will be equipped at first light with a multi-conjugate adaptive optics system that can provide a high-quality correction over a field of view of over 2 arcmin and feed the corrected light into up to three instruments. The adaptive optics system increases the sensitivity or reduces the time required to reach a certain signal to noise by up to a factor of about 200 compared to existing 8 m telescopes.

The TMT has been designed to be a general-purpose observatory, capable of addressing a wide range of science programmes, not focused on a set of specific science goals. A range of instrument types was considered during the design of the observatory; this approach ensured the versatile, general-purpose nature of the observatory (Figure 6). The lifetime of the observatory is such that the telescope will be required to support instruments that we have not yet envisioned and so all attempts have been made to design a telescope and enclosure that will be able to support multiple forms of instruments. The first light instruments that have been chosen and included in the observatory construction plans are versatile workhorse instruments, capable of supporting many of the science programmes that have been envisioned for the TMT.

*The 30 m diameter primary mirror and 450 m effective focal length*

The 30 m diameter primary mirror gives the enhanced light-gathering power required for the envisioned science and the 450 m effective focal length gives a plate scale at the focal plane that matches the diffraction limit of the telescope to the size of pixels in optical and infrared detectors. The full field of view of the telescope at the instrument focal stations on the Nasmyth platforms is 20 arcmin.

*Adaptive optics systems and high-quality telescope optics*

For adaptive optics to work effectively and give high-quality corrections, the telescope optics have to be of very high quality. The most stringent case is when observing faint exoplanets that are close to their host star, namely high-contrast extreme adaptive optics observations. The adaptive optics system can correct for almost all of the atmospheric turbulence, allowing the bright host star light to be masked out so that the faint planet can be seen. However, any imperfections with the telescope optics will scatter some of the light from the host star into the region where the faint planet resides, reducing the achievable contrast or the ability to detect the planet against the background light. The requirements on segment polishing and the performance of the primary mirror control system are set to support extreme adaptive optics observations.

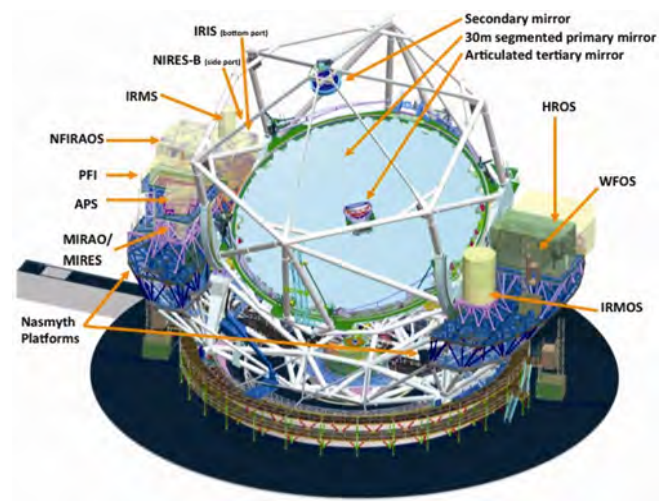

**Figure 6.** Conceptual instrument suite located on the Nasmyth platforms of the telescope. The first light configuration includes IRIS, WFOS, IRMS and NFIRAOS.

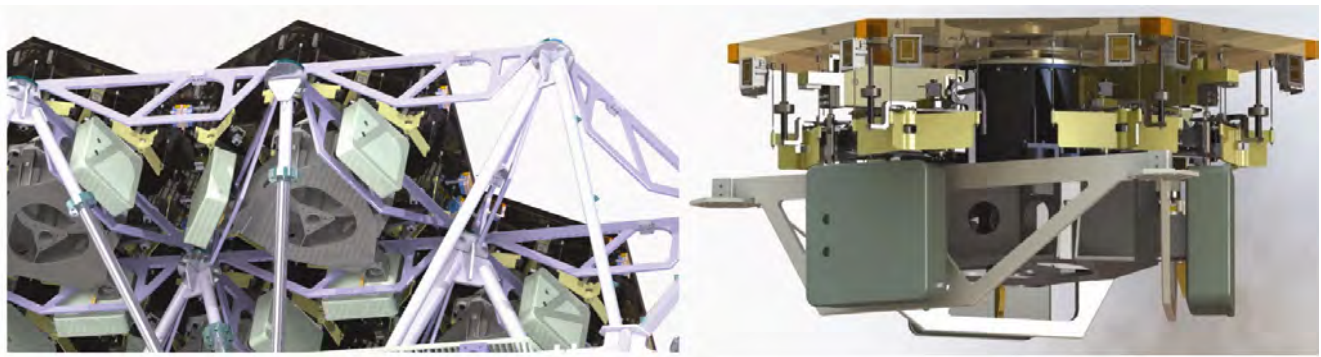

**Figure 7.** (Left) View of the underside of the primary mirror, showing the mirror segments and the support structures and control system. (Right) Green boxes are the M1 segment actuators that hold the mirror segments in the correct position with respect to each other. The segment edge sensors are shown (gold-coloured blocks on the underside edges of the segment) and gold-coloured whiffletree system for segment warping is also seen.

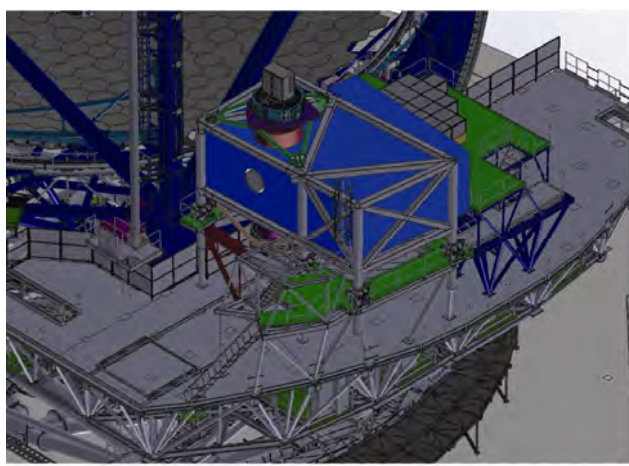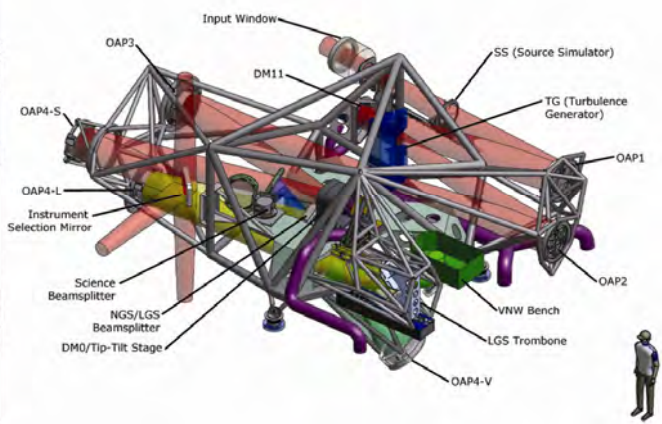

**Figure 8.** (Left) Blue structure is the NFIRAOS adaptive optics system on one of the Nasmyth platforms of the TMT. (Right) Internal design of NFIRAOS. Orientation is the same as in the image on the left. The three potential instrument output ports are seen.

Primary mirror segment polishing errors can be corrected using a whiffletree system for warping the mirror segment and all segments are held in position by the segment actuators that respond to measurements from the segment edge sensors (Figure 7).

*Rapid response*

To study many of the forms of transient objects, it is necessary for the telescope and instruments to respond rapidly when an alert is received. The requirements for the maximum response time from the moment that rapid response or target of opportunity (ToO) observations are triggered is 5 min if using the same instrument, or 10 min if an instrument change is required. This is the total time needed for telescope slewing and all instrument set up until the first science exposure is started.

*Observing modes*

A range of observing modes is being planned for the TMT to cater for the differing needs of Principal Investigators (PIs) across the partnership. It is planned that the majority of observations will be carried out as queue scheduled observations, following a predetermined procedure developed by the PI. Eavesdropping modes will be possible and remote PI-led observing with quick-look data products allowing real time decision-making. Large cross-partner programmes are also anticipated to be in demand and are a way that many astronomers can be supported with a well-defined, systematically obtained suite of observations. Small programmes and classical observing will also be supported.

**Table 1.** The planned TMT instrumentation suite. The first three instruments have been selected for delivery as part of the TMT construction project and others are potential future instruments at this time, but were considered when designing the observatory

| Instrument | $\lambda$ (μm) | Field of view/ slit length | Spectral resolution | Science cases |
|---|---|---|---|---|
| InfraRed Imager and Spectrometer (IRIS) | 0.8–2.5<br>0.6–5 (goal) | <3″ IFU<br>>15″ imaging | >3500<br>5–100 (imaging) | • Assembly of galaxies at high $z$<br>• Black holes/AGNs/Galactic Center<br>• Resolved stellar populations in crowded fields |
| Wide-field Optical spectrometer and imager (WFOS) | 0.31–1.0 | >40 arcmin$^2$<br>>100 arcmin$^2$ (goal)<br>slit length >500″ | 1000–5000@ 0.75″ slit<br>>7500@0.75″ (goal) | • IGM structure and composition at $2 < z < 6$<br>• Stellar populations, chemistry and energetics of $z > 1.5$ galaxies |
| InfraRed Multislit Spectrometer (IRMS) | 0.95–2.45 | 2 arcmin field, up to 120″ total slit length with 46 deployable slits | $R = 4660$@0.16 arcsec slit | • Early Light<br>• Epoch of peak galaxy building<br>• JWST follow-ups |
| Deployable, multi-IFU, near-IR spectrometer (IRMOS) | 0.8–2.5 | 3″ IFUs over >5′ diameter field | 2000–10,000 | • Early Light<br>• Epoch of peak galaxy building<br>• JWST follow-ups |
| Mid-IR AO-fed Echelle spectrometer (MIRES) | 8–18<br>4.5–28 (goal) | 3″ slit length<br>10″ imaging | 5000–100,000 | • Origin of stellar masses<br>• Accretion and outflows around protostars<br>• Evolution of gas in protoplanetary disks |
| Planet Formation Instrument (PFI) | 1–2.5<br>1–5 (goal) | 1″ outer working angle,<br>0″.05 inner working | $R \leq 100$ | • 10$^8$ contrast ratio (10$^9$ goal)<br>• Direct detection and spectroscopic characterization of exoplanets |
| Near-IR AO-fed echelle spectrometer (NIRES) | 1–5 | 2″ slit length | 20,000–100,000 | • IGM at $z > 7$, gamma-ray bursts<br>• Local Group abundances<br>• Abundances, chemistry and kinematics of stars and planet-forming disks<br>• Doppler detection of terrestrial planets around low-mass stars |
| High-resolution optical spectrometer (HROS) | 0.31–1.1 | 5″ slit length | 50000 | • Doppler searches for exoplanets<br>• Stellar abundance studies in Local Group<br>• ISM abundance/kinematics<br>• IGM characteristics to $z \sim 6$ |
| 'Wide'-field AO imager (WIRC) | 0.8–5.0 | 30″ imaging field | 5–100 | • Precision astrometry (e.g. galactic center)<br>• Resolved stellar populations out to 10 Mpc |

*First light and future instruments*

All instruments on the TMT will be located on the Nasmyth platforms (Figure 6), to provide a constant gravity vector and help significantly with instrument stability. Multiple instruments will be ready and available at any time, and can be selected by rotating the tertiary mirror. A subset of initial instrument ideas were selected for continued development as part of the observatory construction project and is planned to be available at first light (Table 1).

*Narrow field infrared adaptive optics system:* Because the majority of science cases for the TMT require spatial sampling at scales below the limits imposed by the atmosphere and many require spatial sampling at scales equal or close to the fundamental diffraction limit of the telescope, a diffraction-limited near-infrared adaptive optics capability has always been planned for first light with the TMT.

The NFIRAOS system (Figure 8), is a multi-conjugate adaptive optics system that gives a high-quality correction over a relatively wide field of 2′ diameter with the highest quality concentrated on the inner 30″ wide field. NFIRAOS has three instrument ports and will provide corrected wavefronts between 0.8 and 2.4 μm.

A laser system can be used to generate false guide stars and increase coverage to almost every position over the sky.

*Infrared imaging spectrograph (IRIS):* The IRIS instrument will be fed from the bottom port of NFIRAOS and provides two parallel and complementary capabilities: (1) a broad band imaging channel with 0.004″ pixels covering a 34″ field of view with 75 different filters, and (2) an integral field spectrograph with spectral resolution of up to $R = 8000$ and selectable field of view/spatial resolution from 0.004″ to 0.05″ per pixel over 0.45″ × 0.51″ to 2.25″ × 4.4″ fields of view (Figure 9).

*Wide field optical spectrograph (WFOS):* A 'workhorse' capability that is needed to support a lot of the science envisioned for the TMT is for a flexible multi-object optical spectroscope with relatively high spectral resolution (up to $R \sim 8000$) for a small number of targets and the ability to observe up to ~200 targets at low spectral resolution over a restricted wavelength range. The design for WFOS provides these capabilities and would use traditional pre-made slit masks (Figure 10).

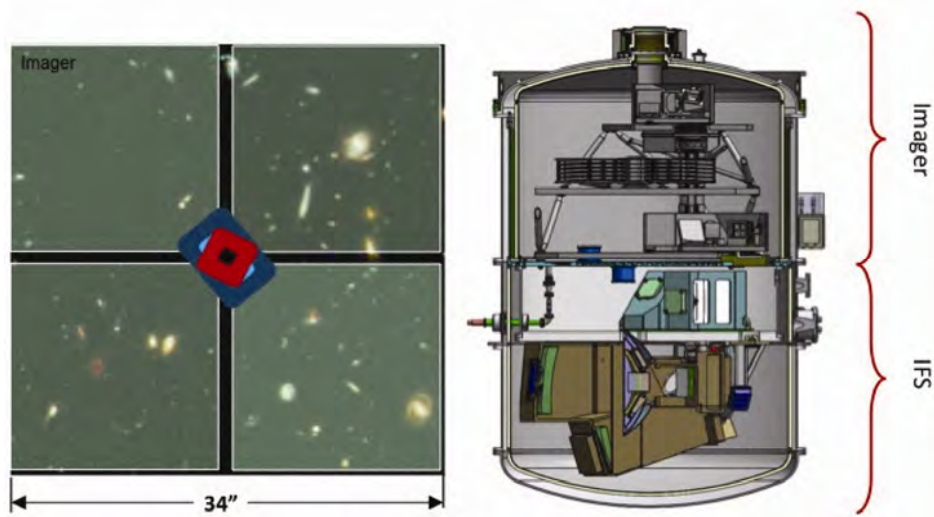

**Figure 9.** (Left) The IRIS field of view. Four 4 k × 4 k detectors are used for the diffraction-limited imager. The coloured rectangles in the centre represent the four options for the field of view of the integral field spectrograph (IFS). (Right) The two-part design of the science instrument with light entering from above. The imager and IFS are being built separately and will be integrated before the instrument is sent to the telescope site.

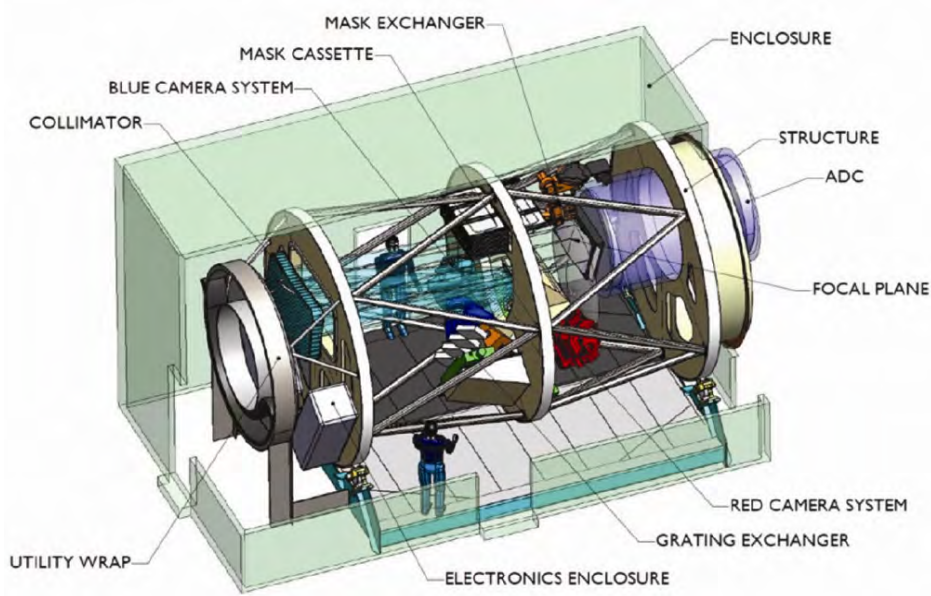

**Figure 10.** An overview of the WFOS design.

Indian scientists and engineers are deeply involved with several aspects of the development of WFOS.

*Infrared multi-object spectrograph (IRMS):* A common capability that is required to carry out a lot of the TMT science is multi-object near-infrared (IR) spectroscopy; however, the target densities and required spatial resolutions are higher than with the optical science cases. A convenient way of providing the necessary near-IR capabilities is to closely copy the successful Keck Observatory instrument MOSFIRE (Figure 11), a multi-object spectrograph that incorporates a reconfigurable cryogenic slit mask that facilitates convenient and efficient changes to the science target distribution. IRMS uses the reconfigurable slit mask and is fed by the NFIRAOS adaptive optics system. IRMS has a pixel scale of 0.06″, far larger than the 0.007″@1 μm diffraction limit, but the field of view is 2′ in diameter. Up to 46 spectra can be simultaneously obtained, or the system can be used for wide and narrow band imaging.

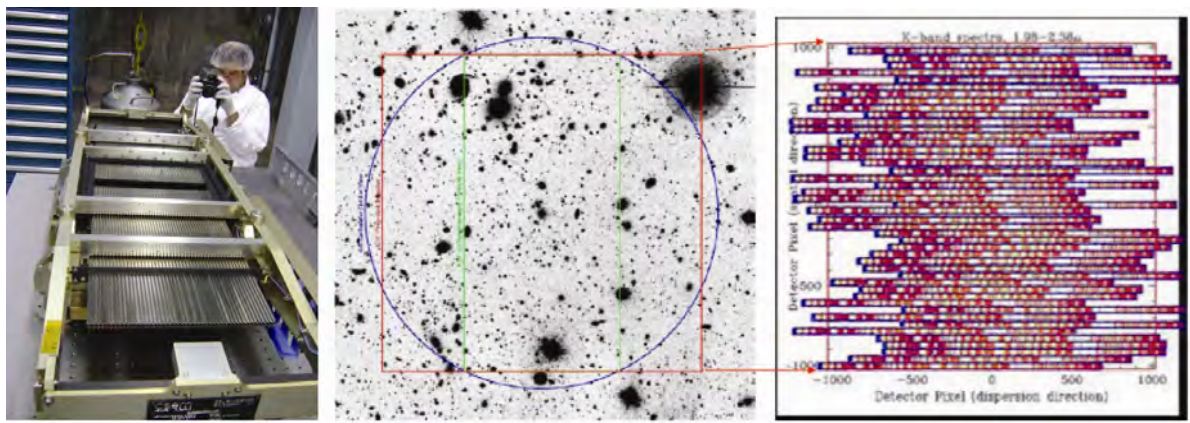

**Figure 11.** (Left) The cryogenic reconfigurable slit mask for MOSFIRE (now used at the Keck Observatory) showing the 46 reconfigurable slits. At the centre is an illustrative image of a field imaged with IRMS. (Right) Regions of the 46 spectra on the detector plane that could be obtained from sources in the field.

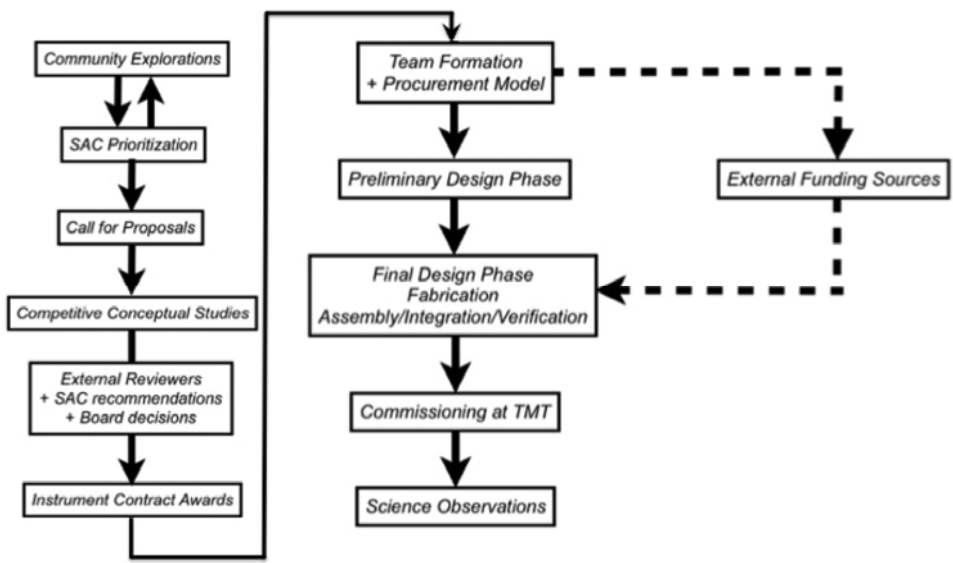

**Figure 12.** Block diagram of the future instrument selection and development process.

*Future instruments and the instrument development process:* A process has been created to identify and develop future instruments with the aim of delivering new capabilities to the observatory every 2–3 years, and that is responsive to community needs and technology readiness (Figure 12). The timescale for instrument development once the call for proposals has been circulated is about 10 years. This depends significantly on the complexity of the instrument and the rate that funding can be secured.

Visitor or private instruments will also be supported, but some time on those instruments will need to be available to the TMT users in a collaborative manner with the visitor instrument team.

**Scientific synergies with other astronomical facilities**

The TMT will enable powerful scientific synergies with a variety of other astronomical facilities. In the case of survey telescopes, the TMT will be used for intensive follow-up observations. With telescopes operating at different wavelengths, the TMT will provide measurements necessary for a comprehensive understanding of the astrophysical phenomenon or objects being studied. A few specific examples of powerful synergies between the TMT and other astronomical facilities that India is involved include:

- Large Synoptic Survey Telescope (LSST) – High redshift/distant quasars, rapid follow-up of transient objects such as supernovae and GRBs, the properties of dark energy and dark matter, populations of stars, variable stars and their use in the cosmic distance ladder, faint dwarf galaxies and local group streams.
- Square Kilometer Array (SKA) – studies of pulsars and tests of general relativity, galaxy evolution, the properties of dark energy and dark matter, the first galaxies and the re-ionization era, signatures of life and transient events.
- Upgraded Giant Meterwave Radio Telescope (uGMRT) – The very first galaxies and the early universe, high redshift radio galaxies and the connection between environment and morphology, emission processes in pulsars, galaxy clusters, faint galaxies within halos, local star formation and the effect of metalliccity, star formation in radio galaxies, quasars and jet/cloud interactions, mysterious transients, galaxy material inflows and outflows and galaxy mergers.
- Laser Interferometer Gravitational wave Observatory (LIGO) – studying the optical counterparts to gravitational wave sources.

## Summary

The TMT observatory will provide new capabilities that will enable significant breakthroughs in many areas of astrophysics. The versatile nature of the observatory means that it can be used to carry out science programmes in areas ranging from studies of solar system objects, to star and planet formation, to characterizing exoplanets and searching for signatures of life, to galaxy evolution, to dark matter and dark energy. We have described how the Observatory has been designed to allow a broad range of discovery space to be probed, and have touched upon the powerful scientific synergies that will be created in concert with other astronomical and scientific facilities that India has leading roles in. The important input in defining the observatory science goals and operations, and designing and delivering the observatory hardware and infrastructure by Indian scientists and engineers is mentioned.

Given the leading role that India is taking with the TMT Observatory and how the TMT will complement other facilities that Indian scientists will utilize, it will be one of the most important scientific facilities for India in the future.

ACKNOWLEDGEMENTS. We acknowledge the support of the TMT collaborating institutions – California Institute of Technology, the University of California, the National Astronomical Observatory of Japan, the National Astronomical Observatories of China and their consortium partners, the Department of Science and Technology of India and their supported institutes, and the National Research Council of Canada. This work was also supported by the Gordon and Betty Moore Foundation, the Canada Foundation for Innovation, the Ontario Ministry of Research and Innovation, the Natural Sciences and Engineering Research Council of Canada, the British Columbia Knowledge Development Fund, the Association of Canadian Universities for Research in Astronomy, the Association of Universities for Research in Astronomy, the US National Science Foundation, the National Institutes of Natural Sciences of Japan, and the Department of Atomic Energy of India. Much of the contents of this manuscript is based on the TMT Detailed Science Case[15], developed by members of the TMT International Science Development Teams. Other contents are based on the India-TMT Science Cases[16].

doi: 10.18520/cs/v113/i04/639-648